\documentclass[prd,showpacs,floatfix,twocolumn,,amsmath,amssymb,floatfix]{revtex4}

\usepackage{graphicx}
\usepackage{epsfig}
\usepackage{epstopdf}
\usepackage{hyperref}
\usepackage{amsmath}
\usepackage{amsfonts}
\usepackage{amssymb}
\usepackage{color}
\begin{document}

\newcommand{\Ima}{\textrm{Im}}
\newcommand{\Rea}{\textrm{Re}}
\newcommand{\mev}{\textrm{ MeV}}
\newcommand{\gev}{\textrm{ GeV}}
\newcommand{\rb}[1]{\raisebox{2.2ex}[0pt]{#1}}

\title{ $B^0 \to D^0 \bar D^0 K^0$, $B^+ \to D^0 \bar D^0 K^+$ and the scalar $D \bar D$ bound state}

\author{L. R. Dai$^{1,2}$}
\email{dailr@lnnu.edu.cn}
\author{Ju-Jun Xie$^{3,4}$}
\email{xiejujun@impcas.ac.cn}
\author{E. Oset$^{2,3}$}
\email{oset@ific.uv.es}
\affiliation{
$^1$ Department of Physics, Liaoning Normal University, Dalian 116029, China\\
$^2$ Departamento de F\'{\i}sica Te\'orica and IFIC, Centro Mixto Universidad \\de Valencia-CSIC, Institutos de Investigaci\'on de Paterna, Apartado 22085, 46071 Valencia, Spain\\
$^3$ Institute of Modern Physics, Chinese Academy of Sciences, Lanzhou 730000, China\\
$^4$ State Key Laboratory of Theoretical Physics, Institute of Theoretical Physics,
Chinese Academy of Sciences, Beijing 100190, China}
\date{\today}

\begin{abstract}

We study the $B^0$ decay to $D^0 \bar D^0 K^0$ based on the chiral unitary  model  that generates the X(3720) resonance, and  make predictions for the $D^0 \bar D^0$ invariant mass distribution.  From the shape of the distribution,  the existence of the resonance below threshold could be induced. We also predict  the rate of  production of the X(3720) resonance to the $D^0 \bar D^0$ mass distribution with no free parameters.

\end{abstract}

\pacs{}

\maketitle

\section{Introduction}

The weak decay of heavy hadrons has brought a wealth of information concerning the interaction of mesons, has reported on many mesonic, as well as baryonic, resonances and, together with devoted theoretical work (see the recent review  \cite{weakreview}), has brought new insight into the nature of many resonances which have been the subject of continuous debate
\cite{Klempt:2007cp,Crede:2008vw}. The meson scalar sector has been emblematic and the chiral unitary approach, unitarizing in coupled channels the information contained in the chiral Lagrangians \cite{Gasser:1983yg}, has shown that the $f_0(500)$, $f_0(980)$, $a_0(980)$ resonances appear as a consequence of the interaction of pseudoscalar mesons and respond to a kind of molecular structure of these components\cite{Oller:1997ti,kaiser,markushin,juanito,rios}, diverting from the standard $q \bar q$ nature of most mesons. The case of the $\sigma$ ($f_0(500)$) has been thoroughly discussed in a recent review \cite{ramonet} and the situation has been much clarified. In this picture, the $f_0(980)$ stands as a bound $K \bar K$ state, with a small component of $\pi \pi$ that provides the decay channel of this state. Much before the advent of the chiral unitary approach, the $K \bar K$ molecular nature of the $f_0(980)$ had already been claimed \cite{Weinstein:1982gc}.  A perspective into these "extraordinary states" was also recently given in the Hadron2015 Conference by Jaffe \cite{jaffe}.

    The chiral Lagrangians can be obtained from a more general framework, which includes vector mesons, the local hidden gauge approach \cite{hidden1,hidden2,hidden3,hidden4}. In this picture the chiral Lagrangians are obtained by exchanging vector mesons between the pseudoscalar mesons.
This picture is most welcome because it allows us to extend the dynamics of the chiral Lagrangians to the heavy quark sector, and the interaction of $D \bar D$, for instance, would be given by the exchange of light vector mesons. Heavy vector mesons could also be exchanged, but their large mass makes the contributions of these terms subdominant, and the dominant terms, where the heavy quarks act as spectators \cite{Xiao:2013yca,Liang:2014eba}, automatically satisfy the rules of heavy quark spin symmetry \cite{Neubert:1993mb,manohar}. It is then not surprising that, in analogy to the $K \bar K$ interaction, which generates the $f_0(980)$, the
$D \bar D$ interaction also gives rise to a bound state, which was studied in \cite{daniel}.
  This state was also predicted in \cite{Nieves:2012tt,HidalgoDuque:2012pq} using effective field theory that implements Heavy Quark Spin Symmetry. In  \cite{Gamermann:2007mu} the results of the $e^+ e^- \to J / \psi D \bar D$ reaction close to threshold \cite{Abe:2007sya} were analyzed. A bump around the $D \bar D$ threshold was observed and the fit to the data was compatible with a state below threshold at 3720 MeV (we shall call this state X(3720) from now on).

     On the other hand, the study of $B$ and $D$ weak decays looking at resonances in the final state, or threshold behavior of invariant mass distributions, has shown that these reactions have a potential to tell us about the existence of "hidden" resonances and their nature. In this sense, in \cite{liang}  a natural explanation was given, in terms of the  $f_0(500)$, $f_0(980)$ as dynamically generated resonances \cite{Oller:1997ti,ramonet}, for the experimental facts that in the $B^0_s \to J/\psi \pi^+ \pi^-$  reaction the  $f_0(980)$ was clearly observed and no trace of the
  $f_0(500)$ was seen \cite{Aaij:2011fx}, while in the case of the $B^0$ decay, the $f_0(500)$
was seen and only a minor fraction of the $f_0(980)$ was observed \cite{Aaij:2013zpt}. These results were complemented by the study of the  $D$ weak decays, were the production of $f_0(500)$, $f_0(980)$ and $a_0(980)$ were studied \cite{dai}.

The idea in this paper will be to make predictions for the $D \bar D$ invariant mass distribution in the decay of $B^0$. In this sense the work of \cite{miguelmari}, where the $B^0_s \to D_s^- (K D)^+$ was studied, showed that from the spectrum of the $KD$ invariant mass one could determine the existence of the $D_{s0}^{\ast\pm}(2317)$ below threshold, and the amount of the $KD$ component in its wave function, using the compositeness sum rule of \cite{danijuan,Hyodo:2011qc,Hyodo:2013nka,Sekihara:2014kya}. A similar work was also done in \cite{liangraquel} where the reactions $\bar B^0 \to \bar K^{*0} X (YZ)$ and $\bar B^0_s \to \phi X (YZ)$ with $X(4160), Y(3940), Z(3930)$ were studied. It was found there that from the study of $D^* \bar D^*$ and $D_s^* \bar D_s^*$ mass distributions close to threshold, the existence of resonances below threshold could be induced.

  In the present paper we will study the $B^0$ decay to $D^0 \bar D^0 K^0$ with a model based on the work of \cite{daniel} that generates the X(3720) resonance, and will make predictions for the $D^0 \bar D^0$ invariant mass distribution. The theory predicts the shape of the distribution close to threshold, but not the absolute normalization. However, from the shape of the distribution the existence of the resonance below threshold could be induced. Additionally, we shall also evaluate the rate of  production of the X(3720) resonance, irrelevant of its decay channel, and will show that the ratio of this rate to the   $D^0 \bar D^0$ mass distribution is then predicted with no free parameters, under the assumption that the X(3720) resonance is dynamically generated. The implementation of the experiment would provide a boost in the search of this elusive state, which we think really exists. This experiment and related ones are currently under investigation by the LHCb collaboration \cite{renato} and this gives us a motivation to perform the calculations at the present time. So far, the related experiment $B^+ \to  D^0 \bar D^0 K^+$ has already been done \cite{Lees:2014abp}. The $D^0 \bar D^0$ invariant mass is measured, but with very small statistics close to threshold. A sharp peak is identified, that  corresponds to the excitation of the $\psi(3770)$ charmonium state, which decays in p-wave in $D^0 \bar D^0$. The X(3720) state is a scalar meson and it decays into $D^0 \bar D^0$ in s-wave. In this sense, testing the invariant mass predicted here should require to separate the s-wave from the p-wave part of the spectrum, something which is already currently been done by the partial wave analysis of the LHCb collaboration, where the contribution of the $\rho$ and $f_0(500)$ are separated in the $B^0$ decay to $J/\psi \pi^+ \pi^-$ \cite{Aaij:2013zpt}. In any case, in the work of \cite{Lees:2014abp}
the contribution of the $\psi(3770)$ is separated and this allows us to make a comparison of our results with this distribution. With present errors we find a good agreement with the data, thus getting extra support for the X(3720) state. However, our study indicates that the
$B^0 \to D^0 \bar D^0 K^0$ reactions is better suited than the $B^+ \to D^0 \bar D^0 K^+$  one to give information on that state.

\section{Formalism}
\begin{figure}[ht]
\begin{center}
\includegraphics[scale=0.6]{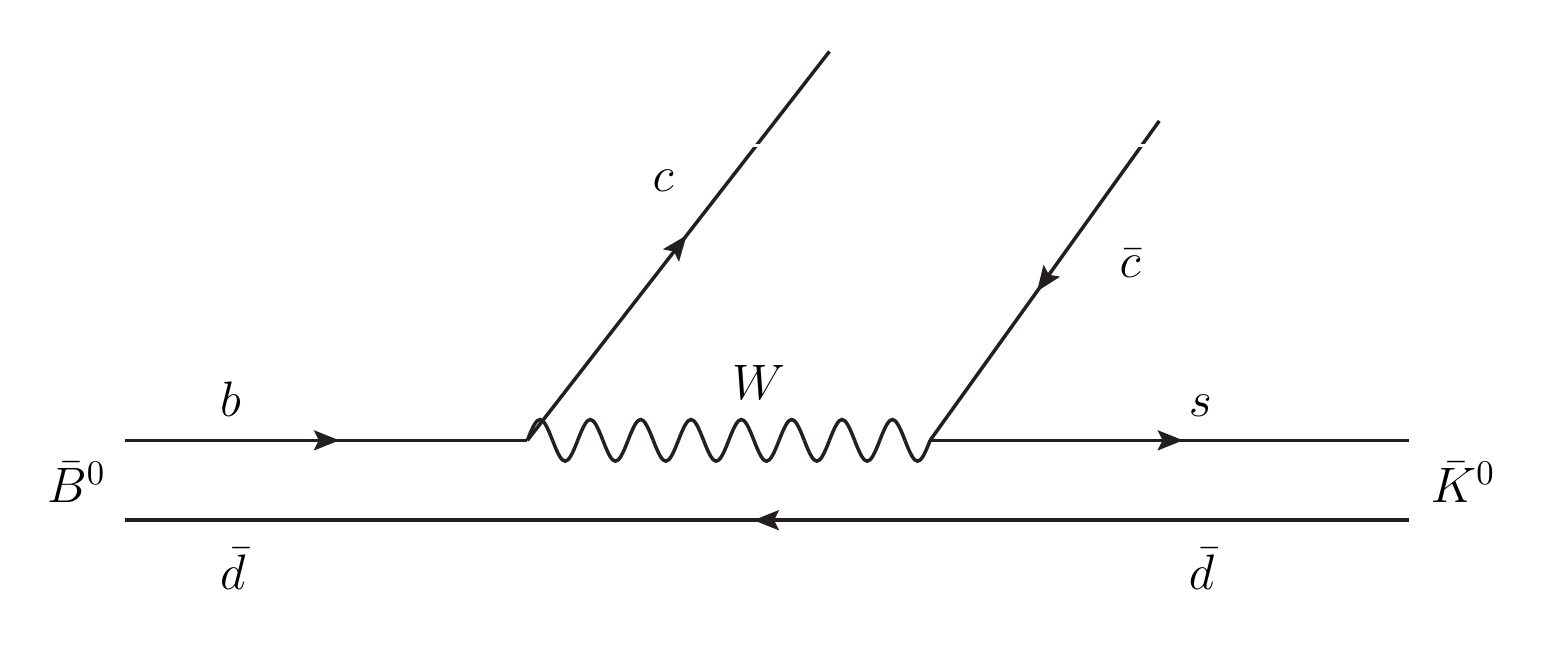}
\caption{Possible diagram at the quark level for  $\bar B^0$ decays into $c\bar c$ and a $s\bar d$ pair.} \label{Fig1}
\end{center}
\end{figure}

\begin{figure}[ht]
\begin{center}
\includegraphics[scale=0.6]{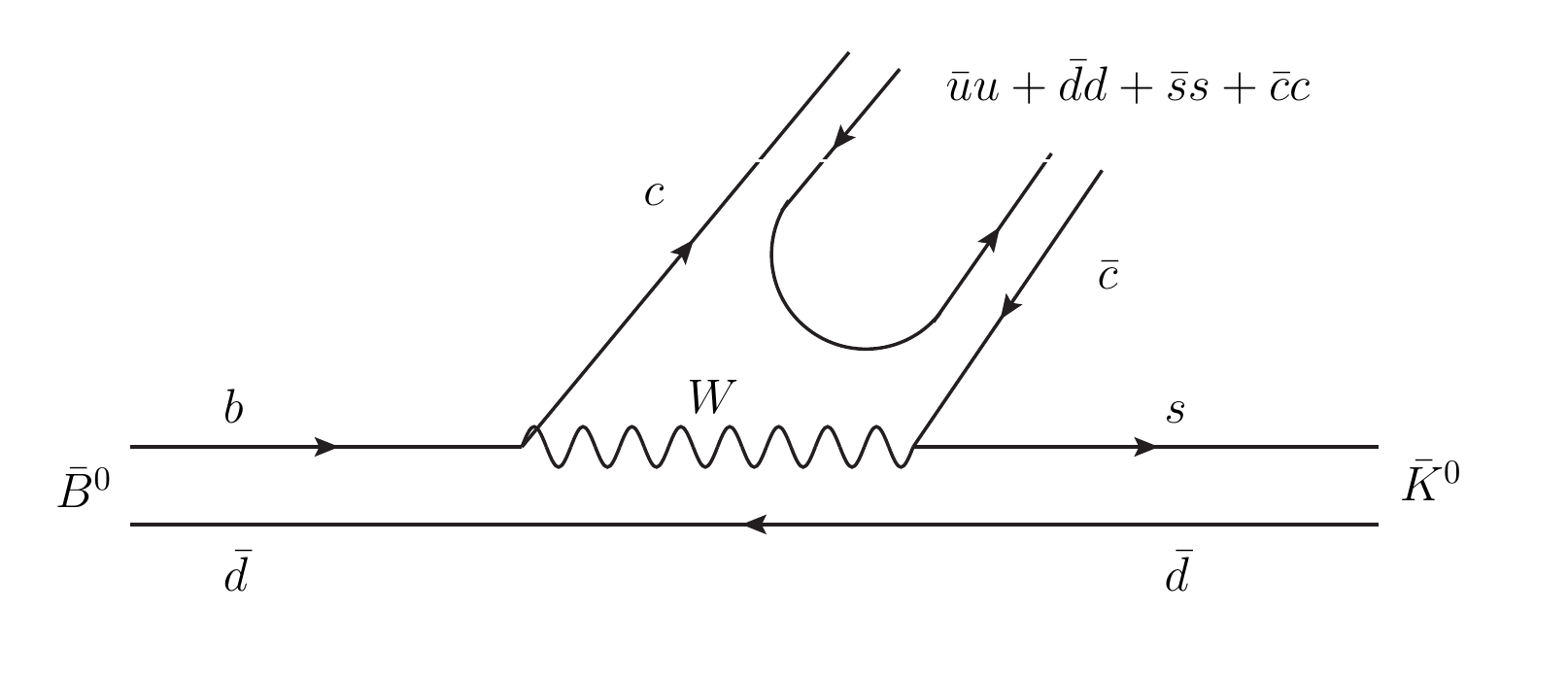}
\caption{Hadronization of the $c\bar c$ pair into two vector mesons for $\bar B^0$ decay.} \label{Fig2}
\end{center}
\end{figure}

If we followed the steps of [26,38], a possible way for the $B^0 \to D^0 \bar D^0 K^0$ to proceed would be  the following: in the first place we would produce a $c\bar c$ together with a $s\bar d$ pair as shown in Fig.\ref{Fig1}.
The next step would consist in introducing a new $q\bar q$ state with the quantum numbers of the vacuum, $\bar uu +\bar d d +\bar s s +\bar c c$ in between the created  $c\bar c$ pair, and
see which combinations of mesons appear. This is depicted in Fig.\ref{Fig2}. This would lead to some $ D \bar D \bar K^0$ components.
\begin{figure}[ht]
\begin{center}
\includegraphics[scale=0.6]{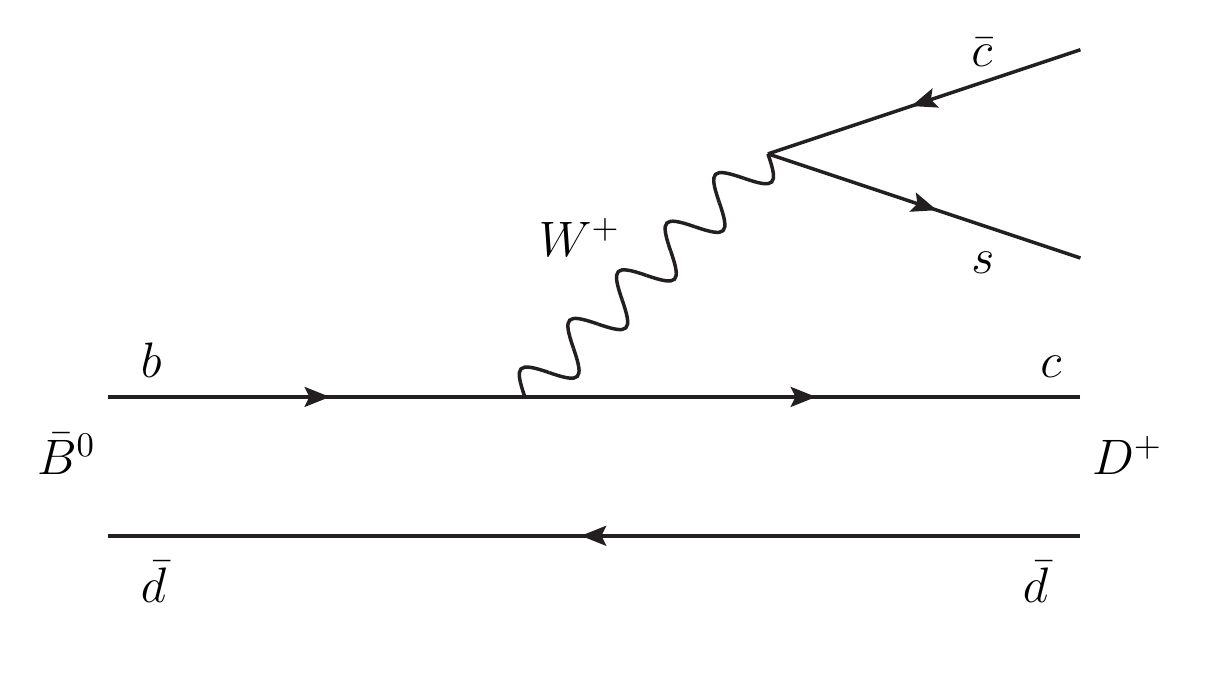}
\caption{Diagram at quark level for external emission of $ s\bar c$ and $ c\bar d$.} \label{Fig3}
\end{center}
\end{figure}
The hadronization is now done including a $ q\bar q$ scalar pair inside the $ s\bar c$ pair, as shown in Fig. \ref{Fig4}.
\begin{figure}[ht]
\begin{center}
\includegraphics[scale=0.6]{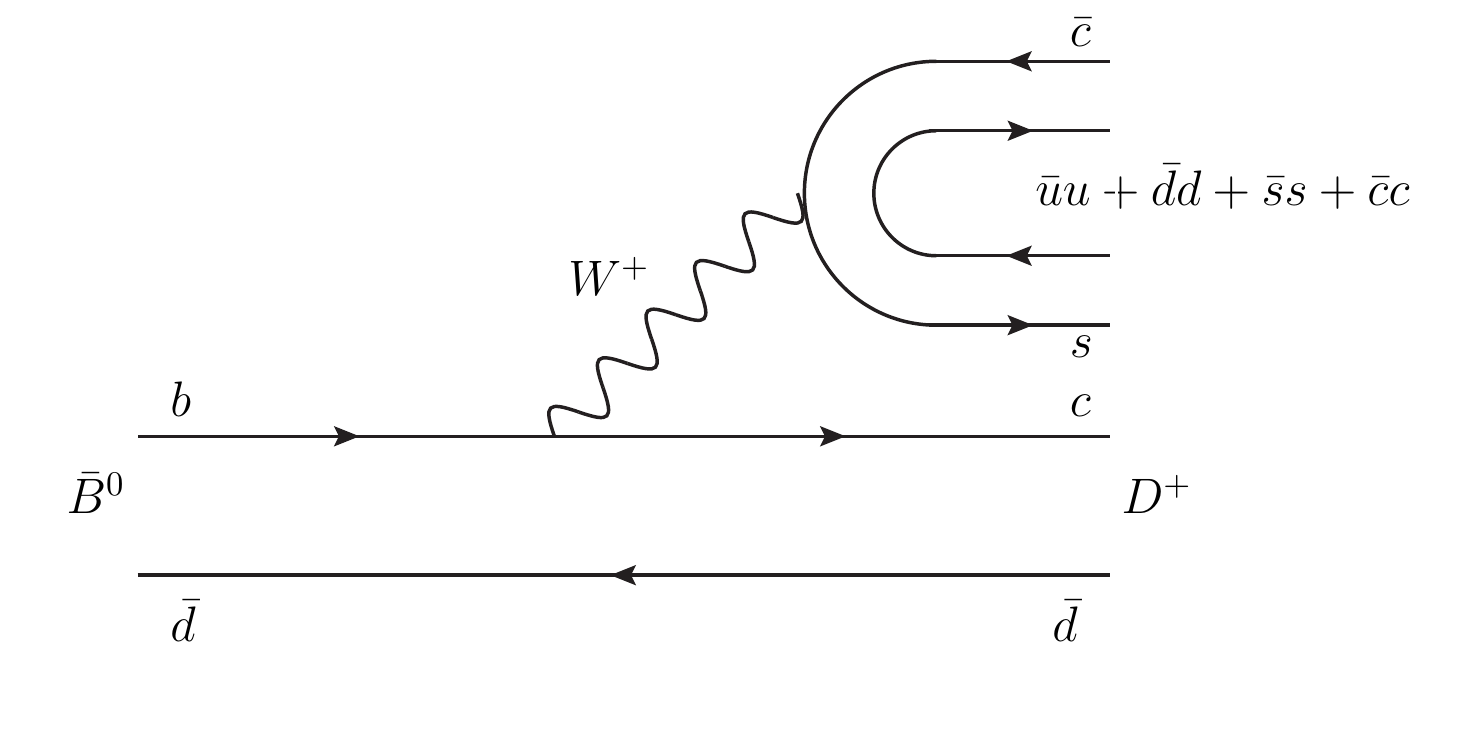}
\caption{The hadronization of the $ s\bar c$ pair into two mesons.} \label{Fig4}
\end{center}
\end{figure}
This way chosen to hadronize the quarks follows the path of \cite{Stone:2013eaa,liang,dai,liangraquel}, based on the topology of the internal
emission \cite{Chau:1982da,Chau:1987tk}. However, in the present case, we also have external emisson, which is color favored and, thus, we adhere to this  other mechanism,
which is depicted in Fig. \ref{Fig3} at the quark level. The hadronization is now done including a $ q\bar q$ scalar pair inside the $ s\bar c$ pair, as shown in Fig. \ref{Fig4},
and technically this is done as follow:
An easy way to see which mesons are produced in the hadronization of $ s\bar c$ is to introduce the $q\bar q$ matrix
\begin{eqnarray}
M = \left(
           \begin{array}{cccc}
             u\bar u & u \bar d & u\bar s & u\bar c\\
             d\bar u & d\bar d & d\bar s & d\bar c\\
             s\bar u & s\bar d & s\bar s & s\bar c\\
             c\bar u & c\bar d & c\bar s & c\bar c\\
           \end{array}
         \right) = \left(
           \begin{array}{c}
            u   \\
             d  \\
             s   \\
             c   \\
           \end{array}
         \right) \left(
           \begin{array}{cccc}
            \bar{u} & \bar{d} & \bar{s} & \bar{c}
           \end{array}   \right).
\end{eqnarray}

\begin{figure}[ht]
\begin{center}
\includegraphics[scale=0.8]{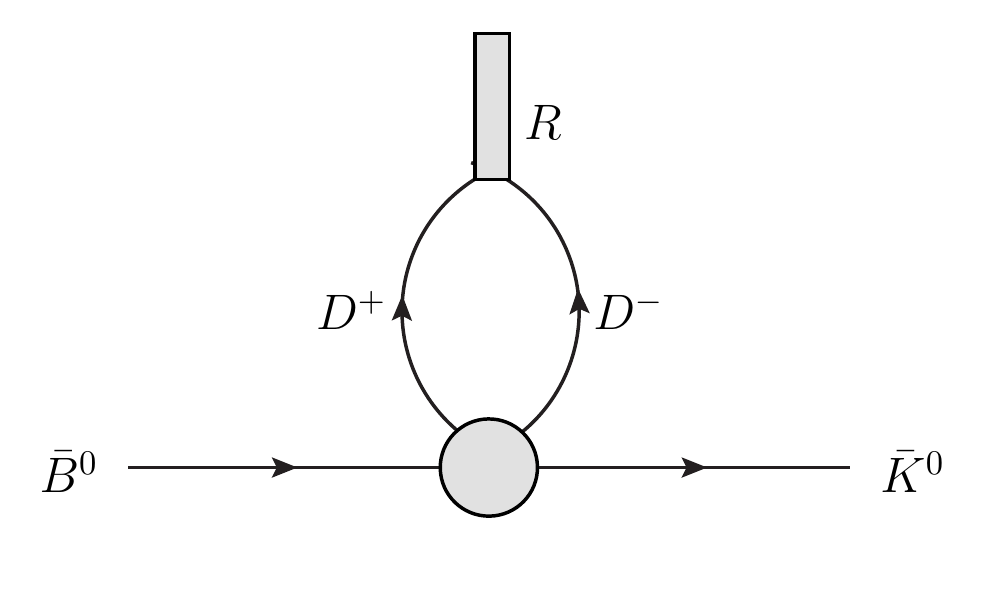}
\caption{Diagrammatic representation of the formation of the resonance $R$ through rescattering of $D^+ D^-$  and coupling to the resonance.} \label{Fig5}
\end{center}
\end{figure}

\begin{figure}[ht]
\begin{center}
\includegraphics[scale=0.8]{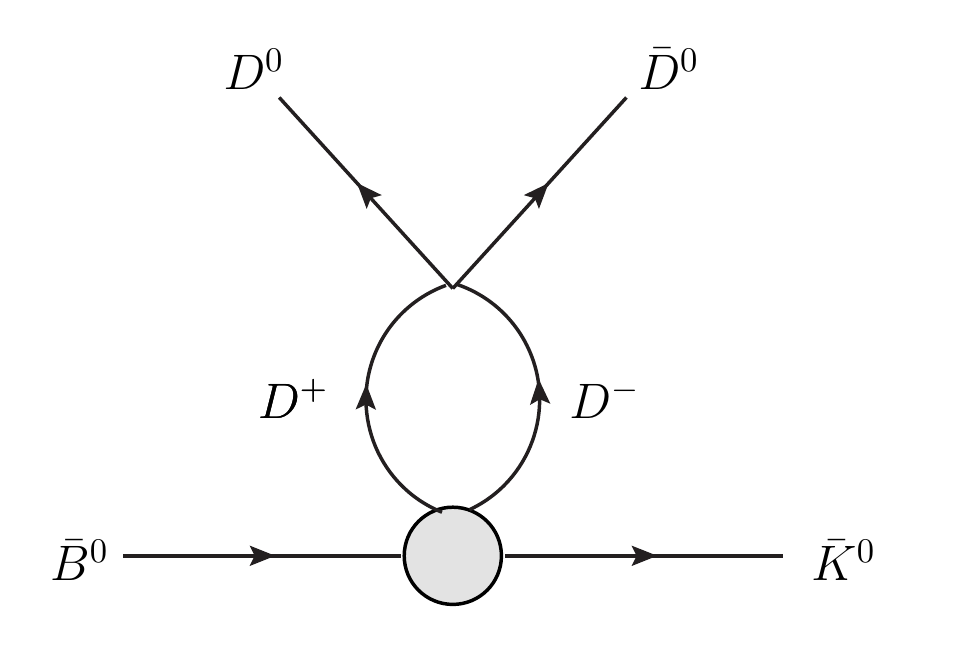}
\caption{Diagrammatic representation of the formation of the resonance $R$ through rescattering of $D^+ D^-$  and coupling to $D^0 \bar D^0$.} \label{Fig6}
\end{center}
\end{figure}

In order to  get the pair of mesons it is convenient to write the  $q\bar q$ in terms of pseudoscalar mesons and then the $M$ matrix
has its equivalent matrix in $\phi$ given by

\begin{equation}\label{eq:Vmatrix}
\renewcommand{\arraystretch}{1.5}
\phi = \left(
           \begin{array}{cccc}
             \frac{\eta}{\sqrt{3}} + \frac{\pi^0}{\sqrt{2}} +  \frac{\eta'}{\sqrt{6}} & \pi^+ & K^{+} & \bar D^{0}\\
             \pi^- &\frac{\eta}{\sqrt{3}} - \frac{\pi^0}{\sqrt{2}} +  \frac{\eta'}{\sqrt{6}}  & K^{0} & D^{-}\\
            K^{-} & \bar{K}^{0} & \sqrt{\frac{2}{3}}\eta'- \frac{\eta}{\sqrt{3}}& D^{-}_s\\
            D^{0} & D^{+} & D^{+}_s & \eta_c\\
           \end{array}
         \right).
\end{equation}
which incorporates the standard $\eta,\eta'$ mixing \cite{Bramon:1992kr}.
Now we see that (see Refs.\cite{miguelmari,MartinezTorres:2009uk})
\begin{eqnarray}
M \cdot M &=& \left(
           \begin{array}{c}
            u   \\
             d  \\
             s   \\
             c \\
           \end{array}
         \right) \left(
           \begin{array}{cccc}
            \bar{u} & \bar{d} & \bar{s} & \bar{c}
           \end{array}   \right)
           \left(
           \begin{array}{c}
            u   \\
             d  \\
             s   \\
             c \\
           \end{array}
         \right) \left(
           \begin{array}{cccc}
            \bar{u} & \bar{d} & \bar{s} & \bar{c}
           \end{array}   \right) \nonumber \\
           &=& \left(
           \begin{array}{c}
            u   \\
             d  \\
             s   \\
             c  \\
           \end{array}
         \right) \left(
           \begin{array}{cccc}
            \bar{u} & \bar{d} & \bar{s}   & \bar{c}
           \end{array}   \right) (\bar{u}u + \bar{d}d + \bar{s}s+ \bar{c}c)
           \nonumber \\
           &=& M (\bar{u}u + \bar{d}d + \bar{s}s + \bar{c}c). \label{MdotM}
\end{eqnarray}
Thus, in terms of mesons the hadronized $ s\bar c$ pair  will be given by
\begin{eqnarray}
&s\bar c (\bar uu +\bar d d +\bar s s + \bar c c) \equiv (M \cdot M)_{34} \equiv (\phi \cdot \phi)_{34} \nonumber\\
=&K^{-}\bar D^{0}+\bar K^{0} D^{-}+(\sqrt{\frac{2}{2}}\eta'-\frac{\eta}{\sqrt{3}})D^-_s +D^-_s\eta_c.  \label{eq:dai}
\end{eqnarray}
We have the meson-meson components of Eq. (\ref{eq:dai}), together with a $D^+$ of Fig. \ref{Fig3}. If we want to have $D \bar D^0$  and $\bar K^0$ at the end,
we must take the $\bar K^0 D^-$ component of Eq. (\ref{eq:dai}) and let the $D^+D^-$ interact to have $D\bar D^0$. Similarly, if we want to
produce the scalar resonance X(3720), which has zero charge, it is also the $\bar K^0 D^-$ component of Eq. (\ref{eq:dai}) that we must take, and we will let
the $D^+D^-$ interact to give the resonance X(3720) at the end.
Hence, diagramatically this latter process is depicted in  Fig. \ref{Fig5}. For the case of $ D^{0} \bar D^{0}$ production the mechanism is depicted in  Fig. \ref{Fig6}.

Analytically we will now have
\begin{equation}\label{eq:amplituT}
t(\bar B^0 \to \bar K^0 R) = V_P G_{D^+ D^-}   g_{R, D^+ D^-},
\end{equation}
where $G_{M_1 M_2}$ is the loop function of the two intermediate meson propagators \cite{Oller:1997ti} and $g_{R, M_1 M_2}$ is the coupling of the resonance to the $M_1 M_2$ meson pair.

Taking into account that, with the doublets $(D^+,-D^0)$ and $(\bar D^0,D^-)$,  the isospin $I=0$ state of $D \bar D$ is
\begin{equation}\label{eq:dai2}
|I=0,D  \bar D \rangle =\frac{1}{\sqrt{2}}( D^+ D^- +D^0\bar D^0),
\end{equation}
eq. (\ref{eq:dai2}) can be rewritten as
\begin{equation}\label{eq:amplituT1}
t(\bar B^0 \to \bar K^0 R) = V_P G_{D^+ D^-}  g^{I=0}_{R, D \bar D} \frac{1}{\sqrt{2}},
\end{equation}
The factor  $V_P$ englobes  the weak amplitudes plus the hadronization factors.  We take it as a constant since we are only concerned about
a restricted range of invariant masses, and the factor prior to the final state interaction
 is smooth in that range \cite{prd89kxw, Daub:2015xja}
 (see section 3.3 of Ref. \cite{weakreview}, for a more complete discussion on this issue and other approaches).

The partial decay width of $\bar{B}^0 \rightarrow \bar{K}^0 R  $ decay will be
\begin{equation}
\Gamma_R=\frac{1}{8\pi}\frac{1}{M^2_{\bar{B}^0}}|t(\bar{B}^0 \rightarrow \bar{K}^0 R)|^2 p_{ \bar{K}^0} \label{eq:width}
\end{equation}
where $p_{\bar K^0}$ is the $\bar K^0$ momentum in the rest frame of the $\bar{B}^0$.

\section{Complementary test of the molecular nature of the resonances}

 In this section we make a test that is linked to the molecular nature of the resonances.
We study the decay $\bar B^0 \to  \bar K^0 D^0 \bar D^0$ close to the  $D \bar D$  threshold depicted in Fig. \ref{Fig6}.

The production matrix  will be given by
\begin{equation}
t(\bar B^0 \to \bar K^0 D^0\bar{D}^0) = V_P G_{D^+D^-} t_{D^+D^- \to D^0\bar{D}^0}
\label{eq:tmat}
\end{equation}
We must evaluate the coupled channels $D^+D^-,D^0\bar{D}^0,D^+_S D^-_s$ amplitudes which will contain $I=0$ and $I=1$, but close to  the  $D \bar D$  threshold they are dominated by $I=0$.  The meson-meson loop function $G$ and the scattering matrix $t_{i \rightarrow j}$ are evaluated following Ref. [21] as discussed later in the results section.

The differential cross section for production will be given by
\begin{equation}
\frac{d\Gamma}{dM_{\rm inv}}=\frac{1}{32\pi^3}\frac{1}{M^2_{\bar{B}^0}} p_{\bar{K}^0}\tilde{p}_D  |t(\bar{B}^0 \rightarrow \bar{K}^0 D^0 \bar{D}^0)|^2 \label{eq:dGama}
\end{equation}
where $p_{\bar K^0}$ is the $\bar K^0$ momentum in the $\bar B^0$  rest frame and $\tilde p_{D}$ the $D$ momentum in the $D \bar D$ rest frame.
 By comparing this equation with Eq. (\ref{eq:width}) for the coalescence production of the resonance in $\bar B^0 \to \bar K^0 ~R$, we find
\begin{eqnarray}
R_{\Gamma}&&=\frac{M^3_R}{p_{\bar{K}^0}\tilde{p}_D} \frac{1}{\Gamma_R} \frac{d\Gamma}{dM_{\rm inv}} \nonumber\\
\nonumber\\&&=\frac{M^3_R}{4\pi^2}\frac{1}{p_{\bar{K}^0}(M_R)}\frac{|t(\bar{B}^0 \rightarrow \bar{K}^0 D^0 \bar{D}^0)|^2 }{|t(\bar{B}^0 \rightarrow \bar{K}^0 R)|^2}  \label{eq:ratio}
\end{eqnarray}
where we have divided the ratio of widths by the phase space factor $p_{\bar{K}^0} \tilde p_{D}$ and multiplied by $M_{R}^3$ to get a constant value at threshold and a dimensionless magnitude.
We apply this method for the X(3720) resonance that couples strongly to $D\bar{D}$.

\begin{figure}[ht]
\begin{center}
\includegraphics[scale=0.7]{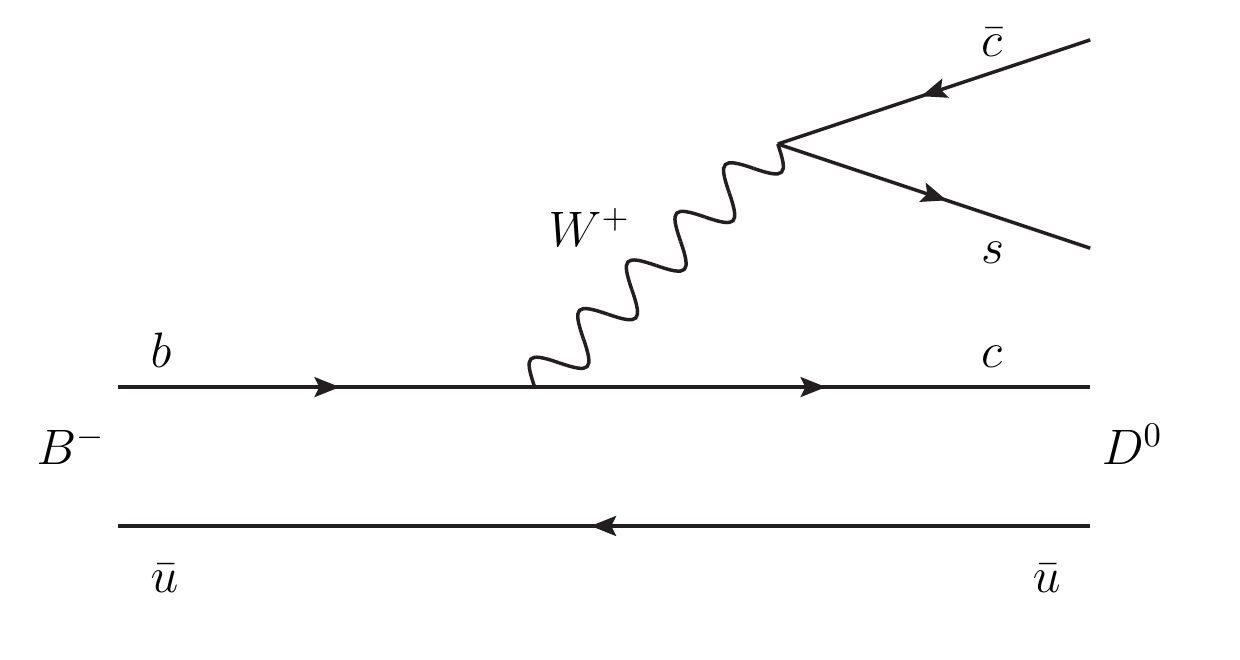}
\caption{The hadronization of the $ s\bar c$ pair into two mesons.} \label{Fig7}
\end{center}
\end{figure}

\begin{figure}[ht]
\begin{center}
\includegraphics[scale=0.65]{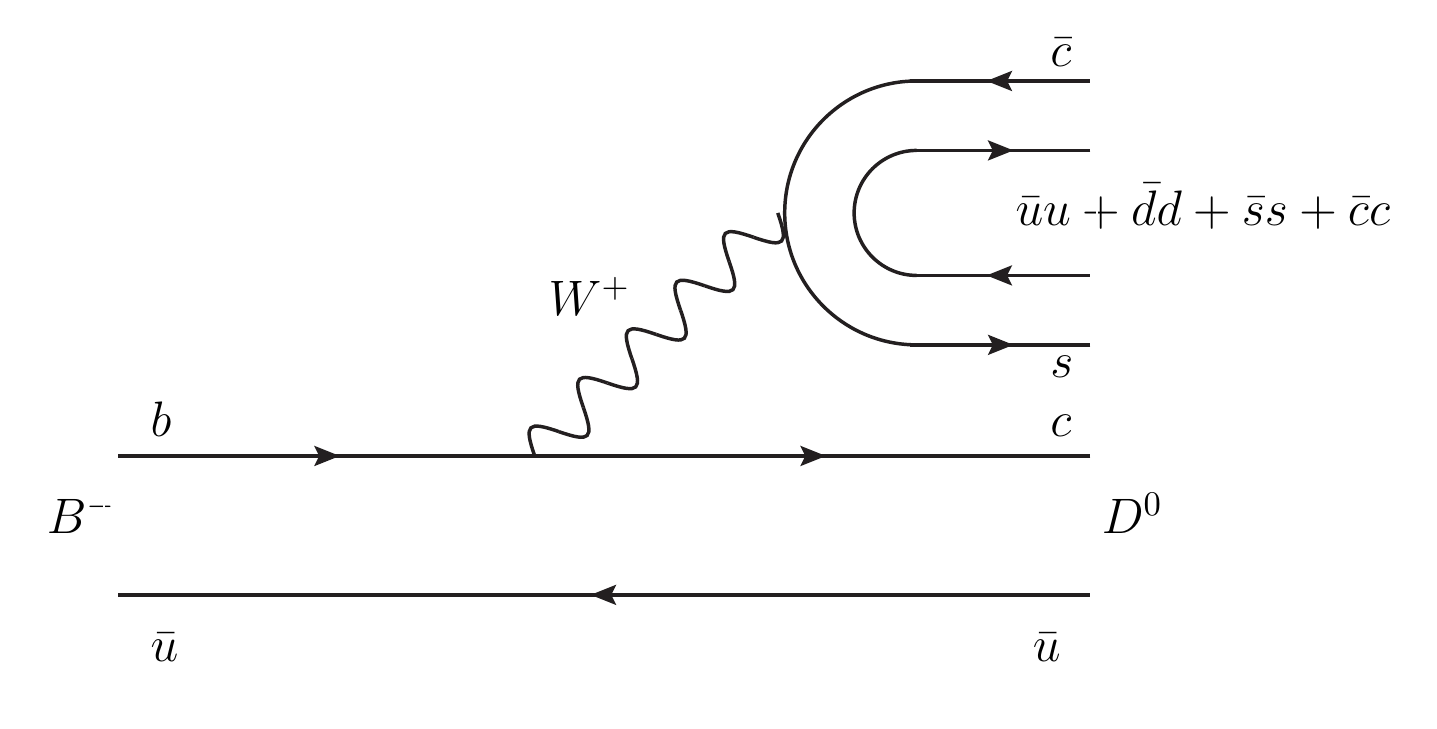}
\caption{The hadronization of the $ s\bar c$ pair into two mesons.} \label{Fig8}
\end{center}
\end{figure}

\begin{figure}[ht]
\begin{center}
\includegraphics[scale=0.7]{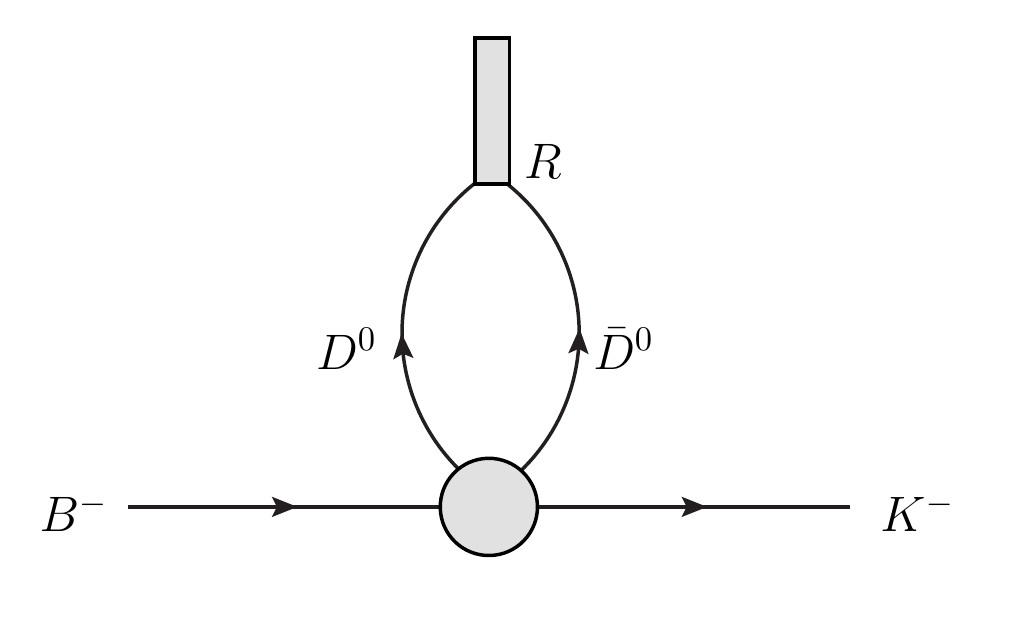}
\caption{Diagrammatic representation of the formation of the resonance $R$ through rescattering of $D^0 \bar D^0$  and coupling to the resonance.}\label{Fig9}
\end{center}
\end{figure}

\begin{figure}[ht]
\begin{center}
\includegraphics[scale=0.5]{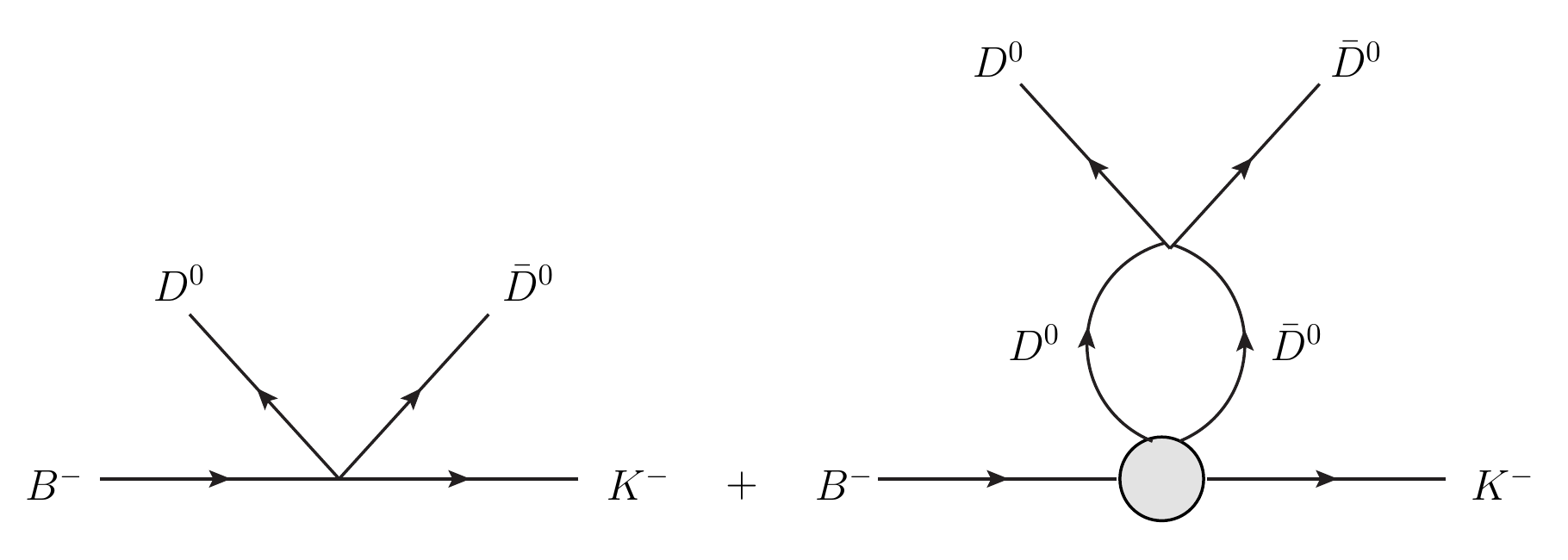}
\caption{Diagrammatic representation of the formation of the resonance $R$ through rescattering of $D^0 \bar D^0$  and coupling to $D^0 \bar D^0$.} \label{Fig10}
\end{center}
\end{figure}

The results obtained are easily translated to the $B^- \to D^0 \bar D^0 K^-$ decay. The diagrams equivalent to Figs. \ref{Fig3},\ref{Fig4},\ref{Fig5},\ref{Fig6},
are now in Figs. \ref{Fig7},\ref{Fig8},\ref{Fig9},\ref{Fig10}. The situation is analogous to  the former one, the hadronization of the $s\bar c $ pair proceeds in the same way  as in  Eq. (\ref{eq:dai}), and the extra $c\bar u $ pair gives rise to a $D^0$, unlike in the former case where the $c\bar u $ pair gave rise to a $D^+ $. Hence, from the  $(K^- \bar D^0 +\bar K^0 D^-)D^0$ contribution at the primary step, we can already produce  $K^- D^0 \bar D^0 $ at tree level and the $D \bar D$ rescattering will be done by the $D^0 \bar D^0$ component.

Then the equation  equivalent to  Eq. (\ref{eq:amplituT1}) and Eq. (\ref{eq:tmat})

\begin{equation}\label{eq:amplituT2}
t(B^- \to K^- R) = V_P G_{ D^0 \bar{D}^0}  g^{I=0}_{R, D \bar D} \frac{1}{\sqrt{2}},
\end{equation}

\begin{equation}
t(B^- \to K^- D^0\bar{D}^0) = V_P(1 + G_{D^0 \bar{D}^0} t_{D^0 \bar{D}^0 \to D^0\bar{D}^0})
\label{eq:tmat2}
\end{equation}
The novelty is in Eq. (\ref{eq:tmat2}) because now we can have $K^-  D^0\bar{D}^0$ production at tree level, and
this is the term unity  in Eq. (\ref{eq:tmat2}).

\section{Results}

First, we use the scattering matrices based on the work of \cite{daniel}, and solve the
Bethe-Salpeter equation in the coupled channels $D^+D^-,D^0\bar{D}^0,D^+_S D^-_s$.

With the $|I=0,D  \bar D \rangle $ wave function of Eq. (\ref{eq:dai2}),
the $I=0$ amplitude for $D  \bar D $ is given by
\begin{equation}
t_{D \bar D \to D \bar D}^{I=0}=\frac{1}{2} (t_{11}+t_{22})+t_{12}
\label{eq:dai3}
\end{equation}
where 1,2,3 stand for the  $D^+D^-,D^0\bar{D}^0,D^+_S D^-_s$ channels. The coupling $g^{I=0}_{R,D \bar{D}}$ is
obtained from this amplitude in the limit
\begin{eqnarray}
t_{D \bar D \to D \bar D}^{I=0} \simeq_{s \to M^2_R} \frac{(g^{I=0}_{R,D \bar{D}})^2}{s-M^2_R}
\label{eq:dai4}
\end{eqnarray}
where $M_R$ is the energy where the pole of the bound $D \bar{D}$ state appears.

The resonance pole  at $\sqrt{s_R}=3719.4+ i 0$ MeV is obtained with no width,  using the same parameters as those in \cite{daniel}, $\alpha=-1.3,\mu=1500$ MeV.

\begin{figure}[htbp]
\begin{center}
\includegraphics[scale=0.36]{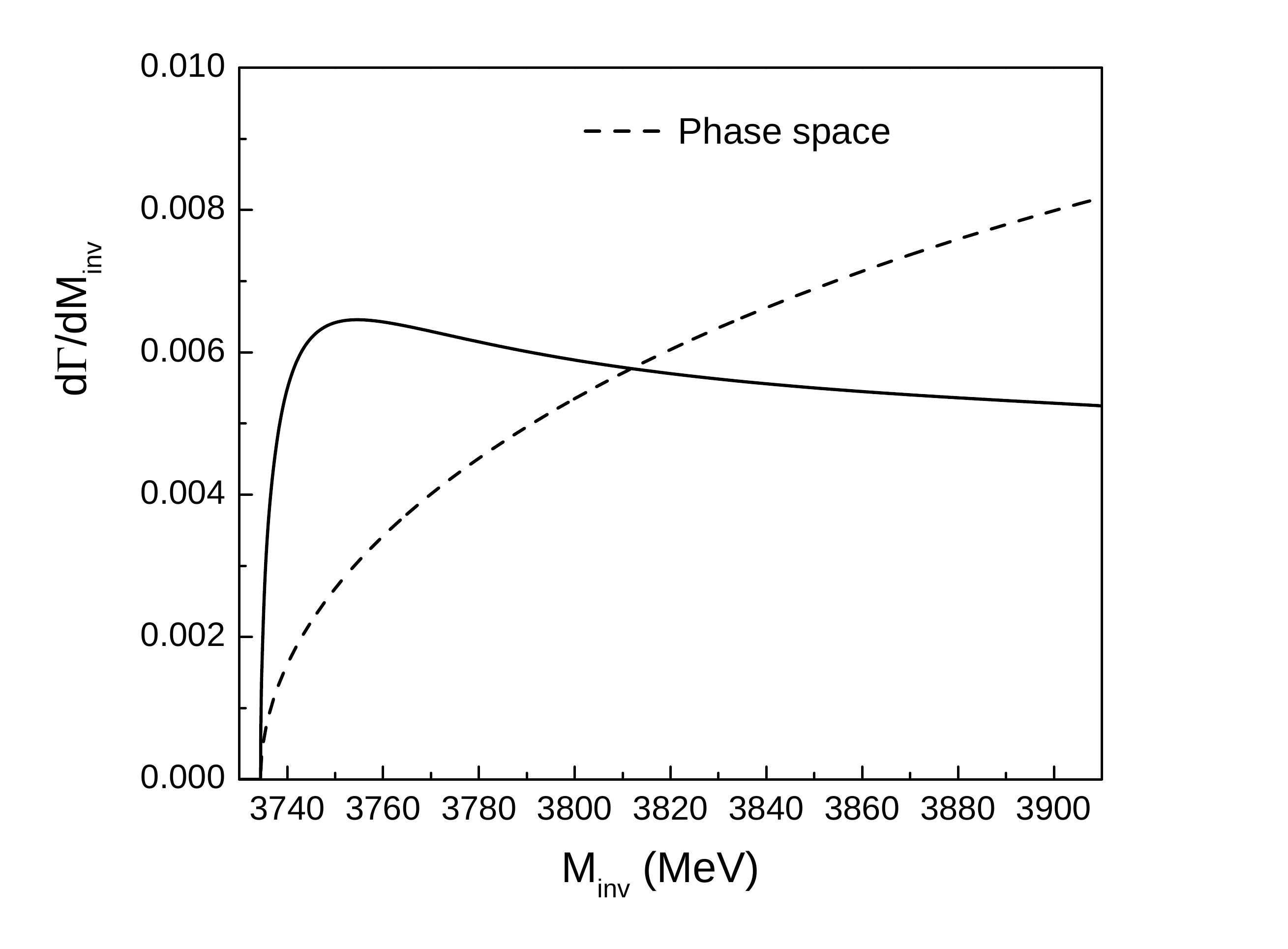}
\caption{The differential cross section for the reaction $\bar B^0 \to \bar K^0 D^0\bar{D}^0$, corresponding to $V_p=1$.
The dashed line corresponds to a phase space distribution normalized to the same area in the range examined.}
\label{dgdm1}
\end{center}
\end{figure}

In the following, from Eq. (\ref{eq:dGama}), we obtain the spectrum  for the $D \bar D$ invariant mass distribution close to threshold in the decay of $\bar  B^0$.
The differential cross section for the reaction $\bar B^0 \to \bar K^0 D^0\bar{D}^0$ is given in Fig. \ref{dgdm1}, where the dashed line corresponds to a phase space distribution normalized to the same area in the range examined.
We can see in the figure that  the shape of the $D^0\bar{D}^0$ mass distribution is quite different from phase space, and this is due to  the presence of the X(3720) resonance below threshold.

\begin{figure}[htbp]
\begin{center}
\includegraphics[scale=0.36]{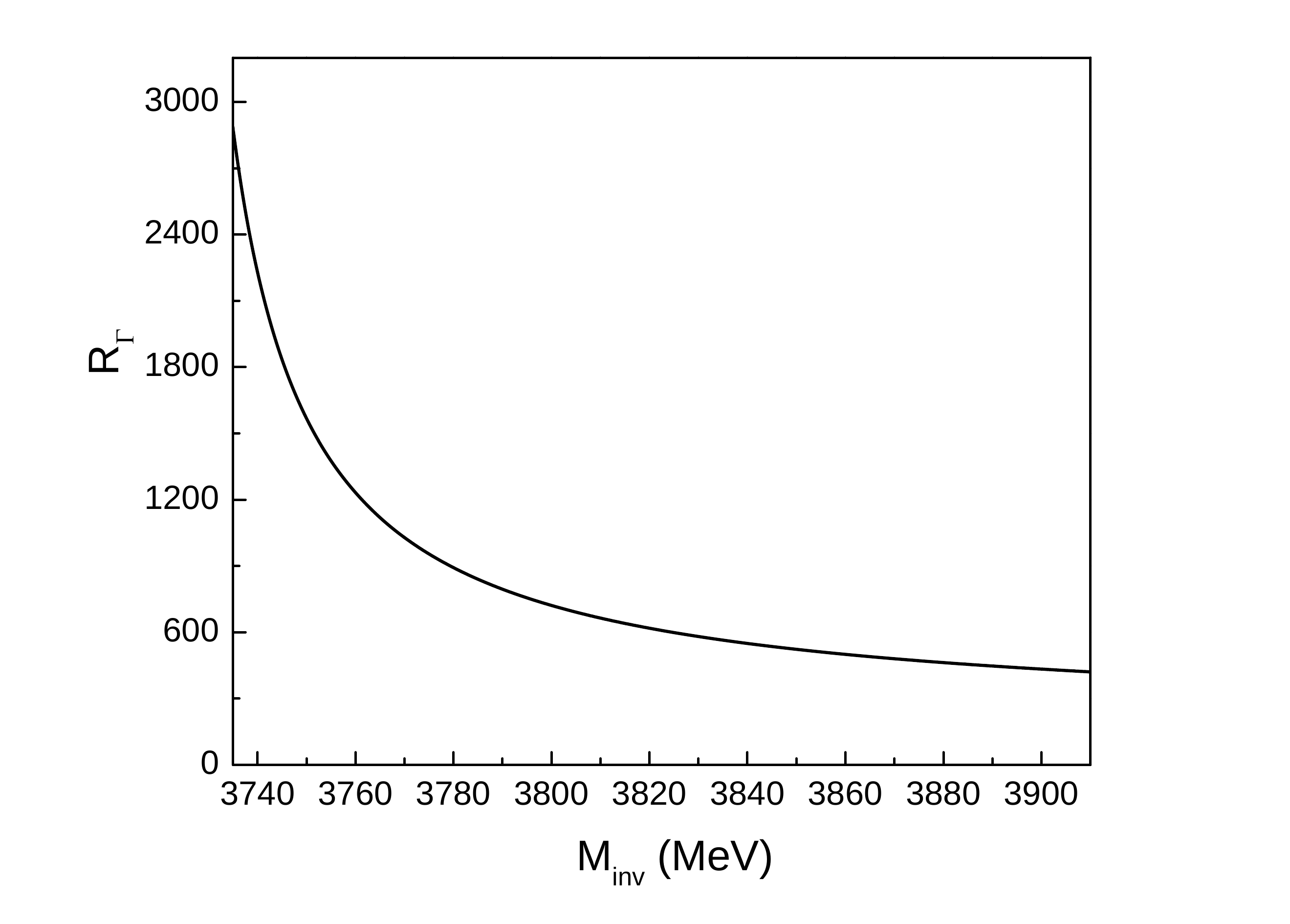}
\caption{Results of $R_{\Gamma}$
of Eq. (\ref{eq:ratio}) as a function of $M_{\rm inv}(D \bar D)$ invariant mass distribution.}
\label{RT1}
\end{center}
\end{figure}

Next have evaluate  Eq. (\ref{eq:ratio}) with the input of Eqs. (\ref{eq:amplituT1})  and (\ref{eq:tmat}), and the results are shown in Fig. (\ref{RT1}).
We observe that the ratio has some structure. There is a fall down of the ratio as a function of energy, as it would correspond to the tail of a resonance below the
threshold of $D \bar D$, the  X(3720), since it basically is giving us the modulus squared of the $t_{D \bar D \to D \bar D}^{I=0}$ amplitude.

Now we turn to the $B^- \to  D^0 \bar D^0 K^-$  reaction. We notice that the related experiment $B^+ \to  D^0 \bar D^0 K^+$ has already been done \cite{Lees:2014abp}, where the $D^0 \bar D^0$ invariant mass is measured, but with very small statistics close to threshold. In the above experiment, a sharp peak is identified which should correspond to the excitation of the $\psi(3770)$ charmonium state, which decays in p-wave in $D^0 \bar D^0$. Since the X(3720) state is a scalar meson,  it couples to $D^0 \bar D^0$ in s-wave.  In any case,  in the work of \cite{Lees:2014abp}, the contribution of the $\psi(3770)$ is separated and this allows us to make a comparison of our results with this distribution.

In order to compare our results with those of Ref. \cite{Lees:2014abp} for $B^+ \to  K^+ D^0\bar{D}^0$, we
subtract from the experimental data the explicit contribution of the $\Psi(3770) $
and $D_{s1}(2700)$ which give some contribution in the region of 100 MeV above $D \bar D$ threshold which we study. We have taken a normalization such as to agree
with that of the data and have collected events  in bins of 40 MeV, integrating $d\Gamma/dM_{\rm inv}$ on the same bins as experiment [threshold,3750],[3750,3790],[3790,3830],[3830,3870],[3870,3910] (units of MeV). The mass distribution $d\Gamma/dM_{\rm inv}(D^0 \bar D^0)$ is now given by
Eq. (\ref{eq:dGama}) substituting $p_{\bar K^0}$ by  $p_{K^-}$ and $t(\bar B^0 \to \bar K^0 D^0\bar{D}^0)$ by $t(B^- \to K^- D^0\bar{D}^0)$
of Eq. (\ref{eq:tmat2}).
We shall come back to this after the discussion of the next section.

\section{Further considerations}
So far we have obtained the X(3720) as a bound $D  \bar D $ state with no width. In practice, this resonance decays into lighter pseudoscalar-pseudoscalar channels as
discussed in Refs. \cite{daniel,xiao13}.  Actually, in Ref. \cite{xiao13} it has found that the width of the X(3720) state decaying to these channels was
$\Gamma=36$ MeV and the most important decay channel was $\eta\eta$. In the present work, we do not want to go through all the coupled channels of \cite{xiao13}
but just wish to have an idea of the effect of considering the width of the  X(3720) state.  For this reason, we work now with four  coupled channels, adding $\eta\eta$
to the former ones, $D^+D^-,D^0\bar{D}^0,D^+_S D^-_s$, and introduce the transition potential $ \eta\eta \to D^+D^-$, $\eta\eta \to D^0\bar{D}^0$ with  a strength
$a$ (dimensionless), similar to that of the model of \cite{xiao13} but tuned to give $\Gamma=36$ MeV just from the $\eta\eta$ decay channel (this is  accomplished with  $a=42$).

\begin{figure}[htbp]
\begin{center}
\includegraphics[scale=0.36]{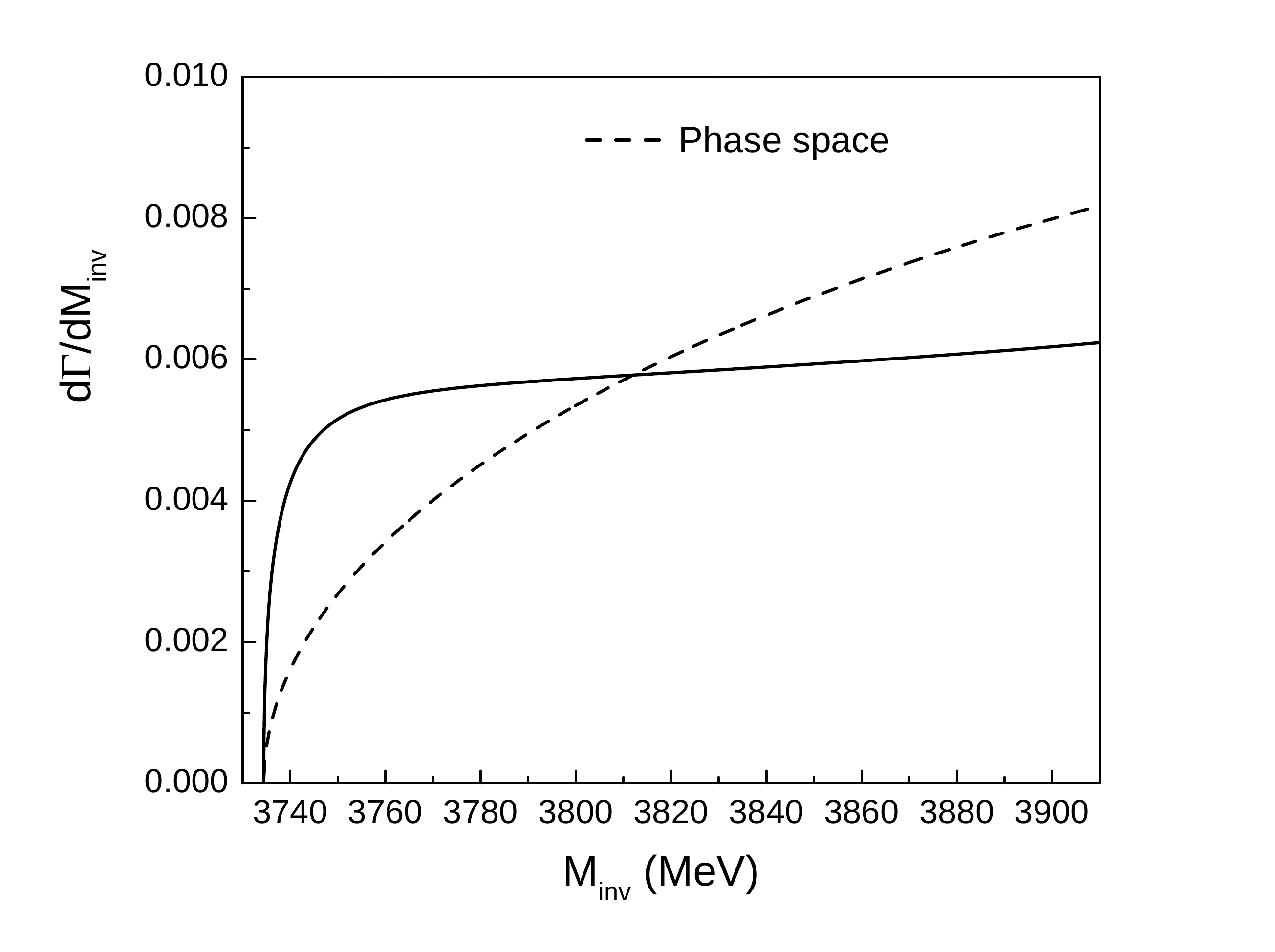}
\caption{The same as those in Fig. (\ref{dgdm1}) but including the $\eta\eta$ channel.
}\label{dgdm1w}
\end{center}
\end{figure}

\begin{figure}[htbp]
\begin{center}
\includegraphics[scale=0.36]{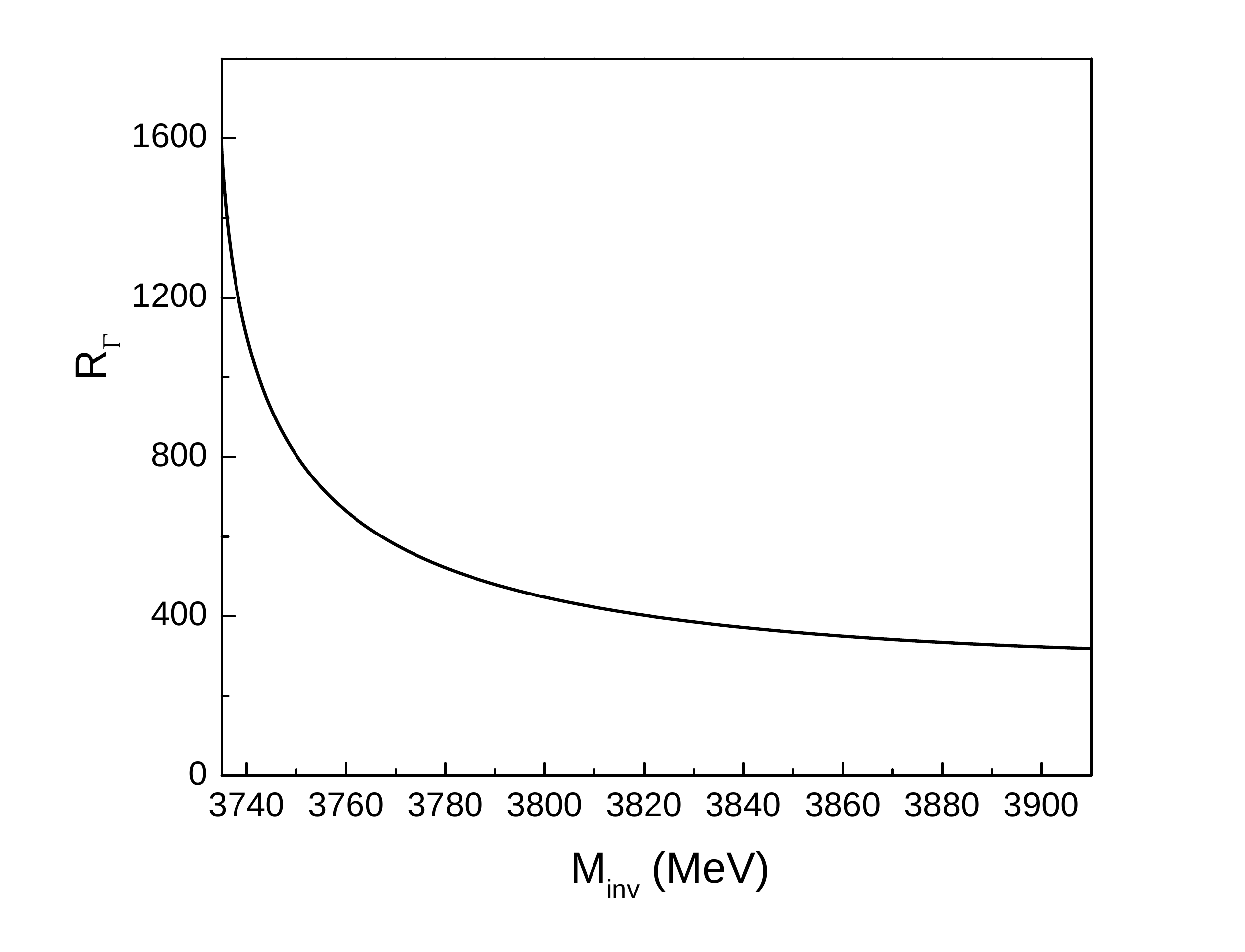}
\caption{The same as those in Fig. (\ref{RT1}) but including the  $\eta\eta$ channel.}
\label{RT1w}
\end{center}
\end{figure}

We perform the same  calculations as before, using the new $t_{D \bar D \to D \bar D}$ amplitude and the results are shown in Figs. \ref{dgdm1w},
\ref{RT1w}. We can see that the features of the mass distribution are very similar to those obtained in the case of zero width for the X(3720) state.
Only the $t_{D \bar D \to D \bar D}$ matrix  becomes  wider in terms of the $D \bar D$ invariant mass and hence the enhancement of the mass distribution
close to the $D \bar D$ threshold is not as strong as before, but clearly is different from a phase space distribution. As a consequence, the ratio of
Eq. (\ref{eq:ratio}) is a bit softer than before, but the fall down in the invariant mass is still clear.

\begin{figure}[htbp]
\begin{center}
\includegraphics[scale=0.36]{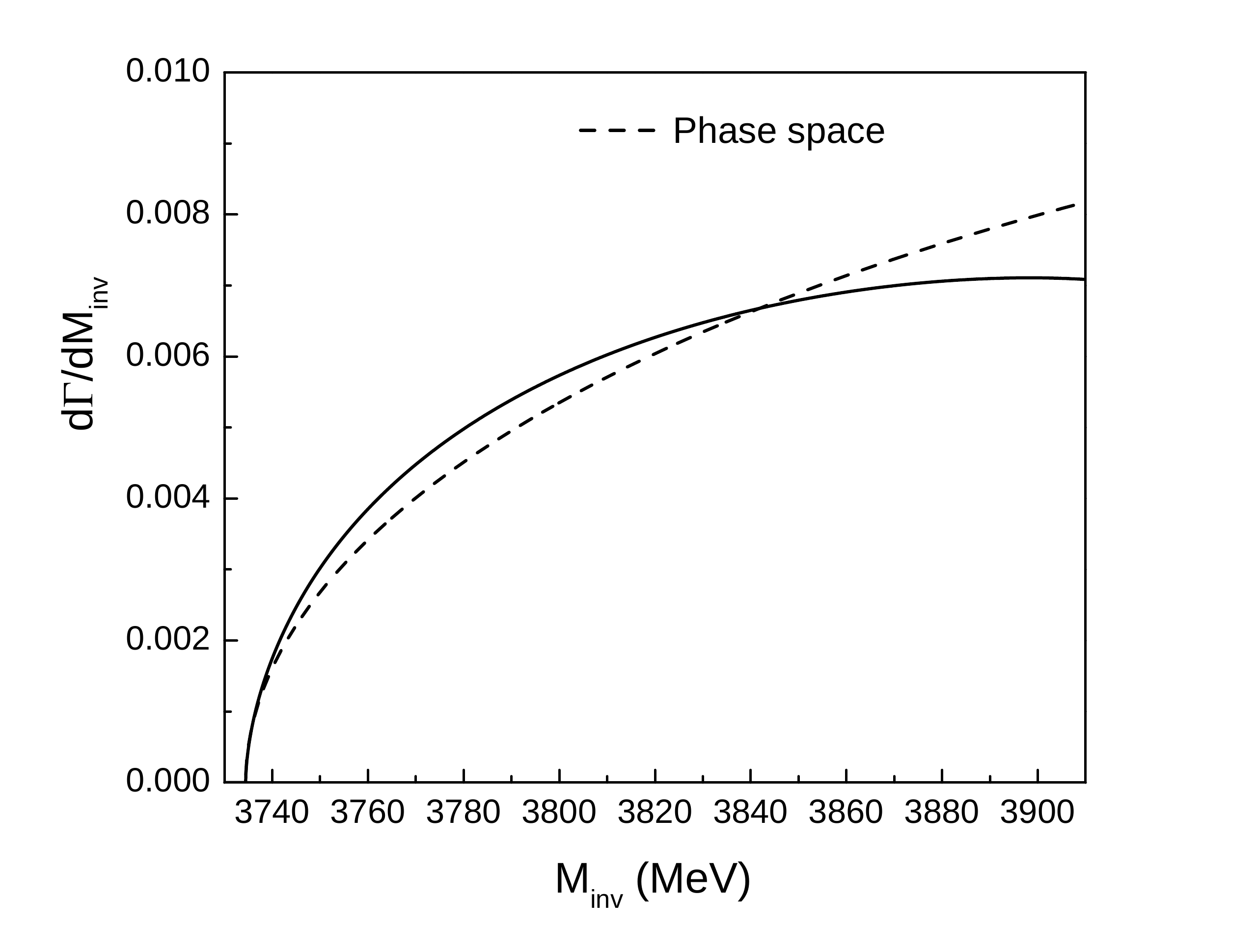}
\caption{The differential cross section for the reaction $ B^- \to  K^- D^0\bar{D}^0$ and
including the $\eta\eta$ channel. The dashed line corresponds to a phase space distribution normalized to the same area in the range examined.}
\label{dgdm2w}
\end{center}
\end{figure}

\begin{figure}[htbp]
\begin{center}
\includegraphics[scale=0.36]{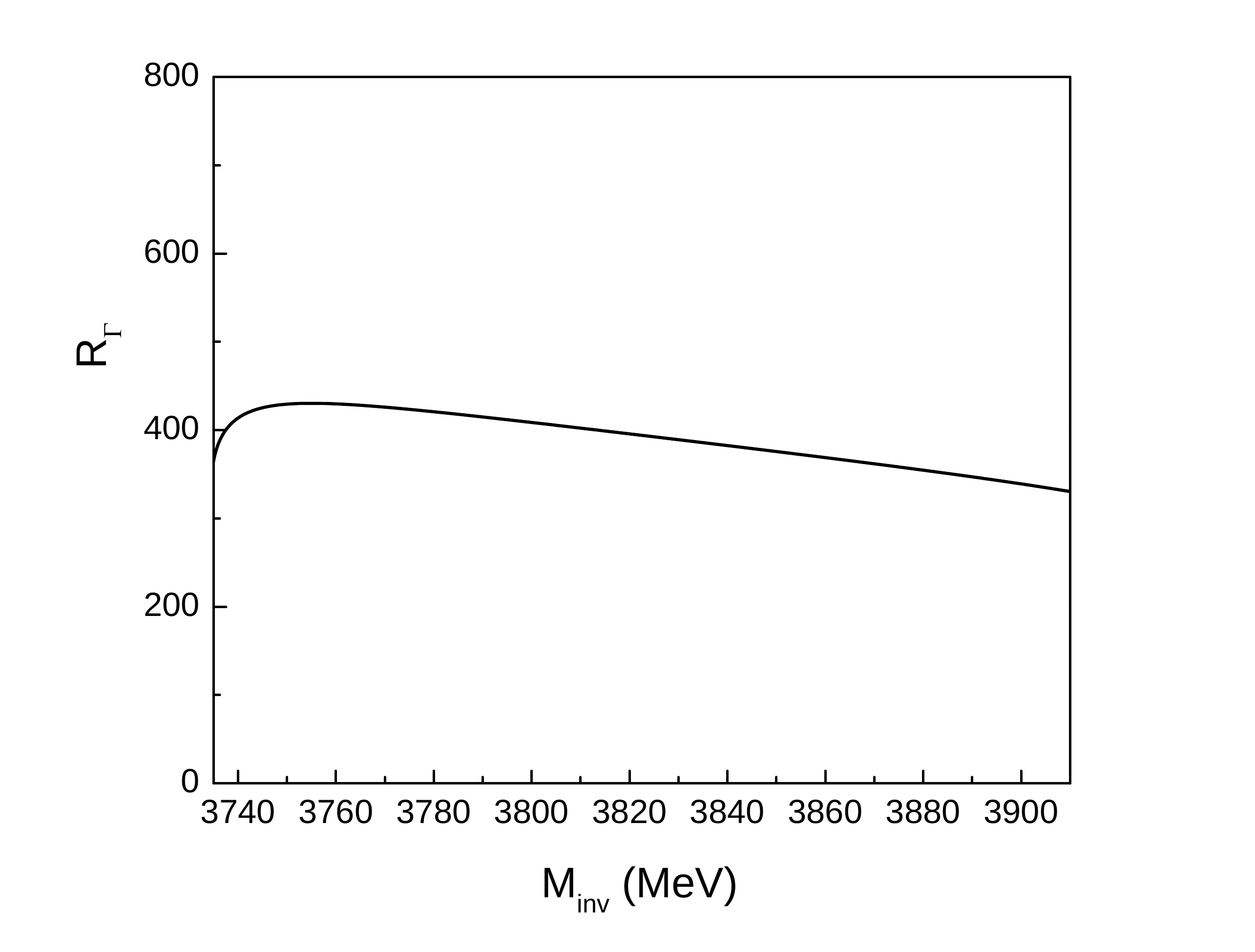}
\caption{Results of $R_{\Gamma}$ of Eq. (\ref{eq:ratio}) as a function of $M_{\rm inv}(D \bar D)$ invariant mass distribution but  for $ B^- \to K^- D^0\bar{D}^0$,
including the $\eta\eta$ channel.}
\label{RT2w}
\end{center}
\end{figure}

\begin{figure}[htbp]
\begin{center}
\includegraphics[scale=0.36]{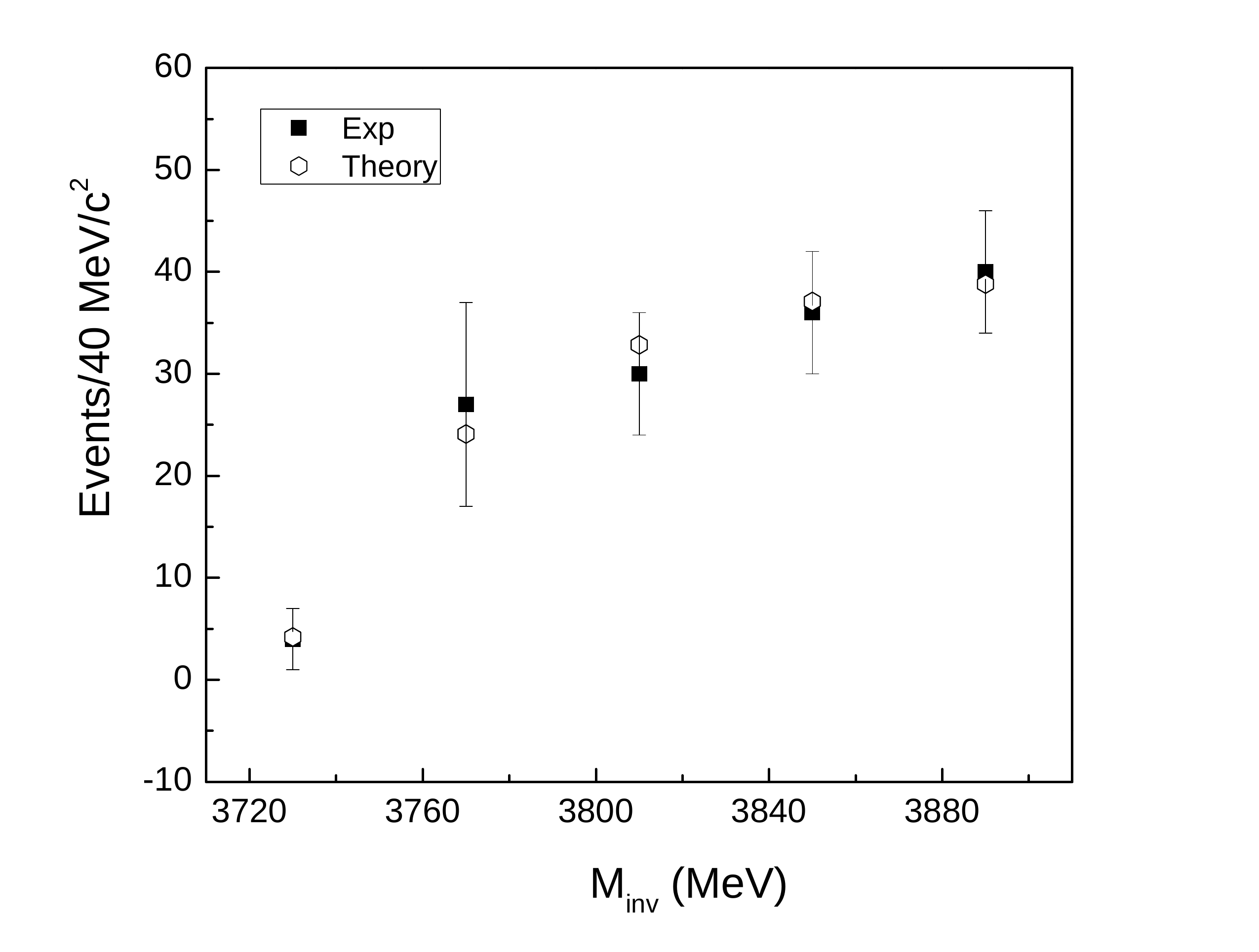}
\caption{Comparison between theory and experiment for the $ B^- \to K^- D^0\bar{D}^0$ decay.}
\label{exp}
\end{center}
\end{figure}

  Finally we now look at the $B^- \to K^- D^0 \bar D^0 $ (which we want to compare with the $B^+ \to K^+ D^0 \bar D^0 $ data)  and we show the results
  in Figs. \ref{dgdm2w},\ref{RT2w},\ref{exp}. In  this case we only  calculate the case with a width for the X(3720).
   In Fig. \ref{dgdm2w} we show $d\Gamma/dM_{\rm inv}$ as a function of the invariant mass. We observe in this case that there is practically
no enhancement close  to threshold and the distribution is closer to phase space. The reason is the term unity in Eq. (\ref{eq:tmat2}), in contrast to Eq. (\ref{eq:tmat}) where only the $t_{D\bar D}$  matrix appears. As a consequence of this,  in Fig. \ref{RT2w}, we do not see the fall down of $R_\Gamma$ that we see in Fig. \ref{RT1w}.
In Fig. \ref{exp} we compare the mass distribution with the data of \cite{Lees:2014abp}.  As mentioned at the end of section IV, we  have
      removed the contributions of $\psi(3770)$ and $D_{s1}(2700)$. Since these contributions come from the analysis of the data of   \cite{Lees:2014abp}, we have
 also taken the results for the total distribution from the same analysis, instead of the raw data.  We normalize the results of Fig. \ref{dgdm2w} to the
 number of  events  in Fig. \ref{exp} and observe that, within errors, the agreement with the data is good. However, we should note that the errors
 are large and the bins of 40 MeV too broad. It would be most helpful to have these results improved with more statistics and better resolution.
      However, the message of the work is that the $B^0 \to D^0 \bar D^0 K^0$ decay is better suited to determine the bound state below threshold, because
      in this reaction we find and justify the presence of an enhancement  of the $D^0 \bar D^0$ mass distribution close to threshold, which is due to the $D\bar D$

The comparison is made here with the limited experimental information available.  Further comparison of these results with coming LHCb measurements will be very valuable to make progress in our understanding of the meson-meson interaction and the nature of the scalar meson  X(3720).

\section{Conclusions}
 In the present paper we have studied the $B^0$ decay to $D^0 \bar D^0 K^0$ based on the chiral unitary  model  that generates the X(3720) resonance, and have made predictions for the $D^0 \bar D^0$ invariant mass distribution.  From the shape of the distribution,  the existence of the resonance below threshold could be induced. Additionally, we have also predicted  the rate of  production of the X(3720) resonance to the $D^0 \bar D^0$ mass distribution with no free parameters,  under the assumption that the X(3720) resonance is dynamically generated.
  So far, the related experiment $B^+ \to  D^0 \bar D^0 K^+$ has already been done, and the $D^0 \bar D^0$ invariant mass is measured, but with very small statistics close to threshold,
   including the s-wave and p-wave parts of the spectrum.  The X(3720) state is a scalar meson and it decays into $D^0 \bar D^0$ in s-wave. In this sense, testing the invariant mass predicted here should require to separate the s-wave from the p-wave part of the spectrum.  In any case, the contribution of the $\psi(3770)$ is separated and this allows us to make a comparison of our results with this distribution.

   With present errors we find a good agreement with the data. However, we found that the
   $B^0$ decay to $D^0 \bar D^0 K^0$ is better suited to study the X(3720) resonance, since
  there is no tree level $D^0 \bar D^0$ production in this decay and this forces  the $D^+ D^- \to D^0 \bar D^0$ transition to intervene to  make $D^0 \bar D^0$ at the end.  The implementation of the experiment in the future would be very helpful  in the search of this elusive state, and  as a further test of the nature  of  the X(3720) resonance.

\section{ACKNOWLEDGMENTS}
 L. R. Dai would like to thank Dr. Z. F. Sun for helpful discussion.
 One of us, E. Oset, wishes to acknowledge support from the Chinese
Academy of Science  in the Program of Visiting Professorship
for Senior International Scientists(Grant No. 2013T2J0012). This work is partly supported
by the Spanish Ministerio de Economia y Competitividad and European
FEDER funds under the contract number FIS2011-28853-C02-01 and
FIS2011-28853-C02-02, and the Generalitat Valenciana in the program
Prometeo, 2009/090. We acknowledge the support of the European
Community-Research Infrastructure Integrating Activity Study of
Strongly Interacting Matter (acronym HadronPhysics3, Grant Agreement
n. 283286) under the Seventh Framework Programme of EU. This work is
also partly supported by the National Natural Science Foundation of
China under Grant Nos. 11575076, 11375080, 11475227,
and supported by Program for Liaoning Excellent
Talents in University under Grant No LR2015032. It is also supported by the Open Project
Program of State Key Laboratory of Theoretical Physics,
Institute of Theoretical Physics, Chinese Academy of Sciences, China (No.Y5KF151CJ1).

\bibliographystyle{ursrt}

\begin{thebibliography}{99}

\bibitem{weakreview}
E. Oset, W. H. Liang, M. Bayar, J. J. Xie, L.R. Dai, M. Albaladejo, M. Nielsen, T. Sekihara, F. Navarra,
L. Roca, M. Mai, J. Nieves, J. M. Dias, A. Feijoo, V. K. Magas, A. Ramos, K. Miyahara, T. Hyodo, D. Jido, M. Doring, R. Molina, H. X. Chen,
E. Wang, L.S. Geng, N. Ikeno, P. Fernandez-Soler, Z. F. Sun, Int. J. Mod. Phys E, in print.

\bibitem{Klempt:2007cp}
  E.~Klempt and A.~Zaitsev,
  Phys.\ Rept.\  {\bf 454}, 1 (2007)
  [arXiv:0708.4016 [hep-ph]].

\bibitem{Crede:2008vw}
  V.~Crede and C.~A.~Meyer,
  Prog.\ Part.\ Nucl.\ Phys.\  {\bf 63}, 74 (2009)
  [arXiv:0812.0600 [hep-ex]].

\bibitem{Gasser:1983yg}
  J.~Gasser and H.~Leutwyler,
  Annals Phys.\  {\bf 158}, 142 (1984).

\bibitem{Oller:1997ti}
  J.~A.~Oller and E.~Oset,
  Nucl.\ Phys.\ A {\bf 620}, 438 (1997)
  [Erratum-ibid.\ A {\bf 652}, 407 (1999)].

\bibitem{kaiser}
N.~Kaiser,
Eur.\ Phys.\ J.\ A {\bf 3}, 307 (1998).


\bibitem{markushin}
M.~P.~Locher, V.~E.~Markushin and H.~Q.~Zheng,
Eur.\ Phys.\ J.\ C {\bf 4}, 317 (1998).

\bibitem{juanito}
J.~Nieves and E.~Ruiz Arriola,
Nucl.\ Phys.\ A {\bf 679}, 57 (2000); Phys.\ Lett.\ B {\bf 455}, 30 (1999).

\bibitem{rios}
J.~R.~Pelaez and G.~Rios,
Phys.\ Rev.\ Lett.\ {\bf 97}, 242002 (2006). 

\bibitem{ramonet}
  J.~R.~Pelaez,
  arXiv:1510.00653 [hep-ph].

\bibitem{Weinstein:1982gc}
  J.~D.~Weinstein and N.~Isgur,
  Phys.\ Rev.\ Lett.\  {\bf 48}, 659 (1982).


\bibitem{jaffe} R. Jaffe in the XVI International Conference on hadron spectroscopy, Jefferson Lab, September, 2015. https://www.jlab.org/conferences/hadron2015/index.html

\bibitem{hidden1}
  M.~Bando, T.~Kugo, S.~Uehara, K.~Yamawaki and T.~Yanagida,
  Phys.\ Rev.\ Lett.\  {\bf 54}, 1215 (1985).

\bibitem{hidden2}
  M.~Bando, T.~Kugo and K.~Yamawaki,
  Phys.\ Rept.\  {\bf 164}, 217 (1988).


\bibitem{hidden3}
  M.~Harada and K.~Yamawaki,
  Phys.\ Rept.\  {\bf 381}, 1 (2003)
  [hep-ph/0302103].

\bibitem{hidden4}
  U.~G.~Meissner,
  Phys.\ Rept.\  {\bf 161}, 213 (1988).

\bibitem{Xiao:2013yca}
  C.~W.~Xiao, J.~Nieves and E.~Oset,
  Phys.\ Rev.\ D {\bf 88}, 056012 (2013)
  [arXiv:1304.5368 [hep-ph]].

\bibitem{Liang:2014eba}
  W.~H.~Liang, C.~W.~Xiao and E.~Oset,
  Phys.\ Rev.\ D {\bf 89}, no. 5, 054023 (2014)
  [arXiv:1401.1441 [hep-ph]].

\bibitem{Neubert:1993mb}
  M.~Neubert,
  Phys.\ Rept.\  {\bf 245}, 259 (1994)
  [hep-ph/9306320].

\bibitem{manohar} A.V. Manohar and M.B. Wise. Heavy Quark Physics, Cambridge Monographs on Particle Physics, Nuclear Physics and Cosmology, vol. 10. Camb. Monogr. Part. Phys. Nucl. Phys.
Cosmol.10,1


\bibitem{daniel}
  D.~Gamermann, E.~Oset, D.~Strottman and M.~J.~Vicente Vacas,
  Phys.\ Rev.\ D {\bf 76}, 074016 (2007)
  [hep-ph/0612179].

\bibitem{Nieves:2012tt}
  J.~Nieves and M.~P.~Valderrama,
  Phys.\ Rev.\ D {\bf 86}, 056004 (2012)
  [arXiv:1204.2790 [hep-ph]].

\bibitem{HidalgoDuque:2012pq}
  C.~Hidalgo-Duque, J.~Nieves and M.~P.~Valderrama,
  Phys.\ Rev.\ D {\bf 87}, no. 7, 076006 (2013)
  [arXiv:1210.5431 [hep-ph]].

\bibitem{Gamermann:2007mu}
  D.~Gamermann and E.~Oset,
  Eur.\ Phys.\ J.\ A {\bf 36}, 189 (2008)
  [arXiv:0712.1758 [hep-ph]].


\bibitem{Abe:2007sya}
  P.~Pakhlov {\it et al.} [Belle Collaboration],
  Phys.\ Rev.\ Lett.\  {\bf 100}, 202001 (2008)
  [arXiv:0708.3812 [hep-ex]].

\bibitem{liang}
  W.~H.~Liang and E.~Oset,
  Phys.\ Lett.\ B {\bf 737}, 70 (2014)
  [arXiv:1406.7228 [hep-ph]].


\bibitem{Aaij:2011fx}
  R.~Aaij {\it et al.} [LHCb Collaboration],
  Phys.\ Lett.\ B {\bf 698}, 115 (2011)
  [arXiv:1102.0206 [hep-ex]].


\bibitem{Aaij:2013zpt}
  R.~Aaij {\it et al.} [LHCb Collaboration],
  Phys.\ Rev.\ D {\bf 87}, no. 5, 052001 (2013)
  [arXiv:1301.5347 [hep-ex]].


\bibitem{dai}
  J.~J.~Xie, L.~R.~Dai and E.~Oset,
  Phys.\ Lett.\ B {\bf 742}, 363 (2015)
  [arXiv:1409.0401 [hep-ph]].

\bibitem{miguelmari}
  M.~Albaladejo, M.~Nielsen and E.~Oset,
  Phys.\ Lett.\ B {\bf 746}, 305 (2015)
  [arXiv:1501.03455 [hep-ph]].


\bibitem{danijuan}
  D.~Gamermann, J.~Nieves, E.~Oset and E.~Ruiz Arriola,
  Phys.\ Rev.\ D {\bf 81}, 014029 (2010)
  [arXiv:0911.4407 [hep-ph]].

\bibitem{Hyodo:2011qc}
  T.~Hyodo, D.~Jido and A.~Hosaka,
  Phys.\ Rev.\ C {\bf 85}, 015201 (2012)
  [arXiv:1108.5524 [nucl-th]].


\bibitem{Hyodo:2013nka}
  T.~Hyodo,
  Int.\ J.\ Mod.\ Phys.\ A {\bf 28}, 1330045 (2013)
  [arXiv:1310.1176 [hep-ph]].


\bibitem{Sekihara:2014kya}
  T.~Sekihara, T.~Hyodo and D.~Jido,
  PTEP {\bf 2015}, 063D04 (2015)
  [arXiv:1411.2308 [hep-ph]].


\bibitem{liangraquel}
  W.~H.~Liang, J.~J.~Xie, E.~Oset, R.~Molina and M.~Döring,
  Eur.\ Phys.\ J.\ A {\bf 51}, no. 5, 58 (2015)
  [arXiv:1502.02932 [hep-ph]].

\bibitem{renato} Renato Quagliani, private communication.

\bibitem{Lees:2014abp}
  J.~P.~Lees {\it et al.} [BaBar Collaboration],
  Phys.\ Rev.\ D {\bf 91}, no. 5, 052002 (2015)
  [arXiv:1412.6751 [hep-ex]].

\bibitem{Stone:2013eaa}
  S.~Stone and L.~Zhang,
  Phys.\ Rev.\ Lett.\  {\bf 111}, no. 6, 062001 (2013).

\bibitem{Bramon:1992kr}
  A.~Bramon, A.~Grau and G.~Pancheri,
  Phys.\ Lett.\ B {\bf 283}, 416 (1992).

\bibitem{Chau:1982da}
  L.~L.~Chau,
  Phys.\ Rept.\  {\bf 95}, 1 (1983).

\bibitem{MartinezTorres:2009uk}
  A.~Martinez Torres, L.~S.~Geng, L.~R.~Dai, B.~X.~Sun, E.~Oset and B.~S.~Zou,
  Phys.\ Lett.\ B {\bf 680} (2009) 310
  doi:10.1016/j.physletb.2009.09.003
  [arXiv:0906.2963 [nucl-th]].


\bibitem{Chau:1987tk}
  L.~L.~Chau and H.~Y.~Cheng,
  Phys.\ Rev.\ D {\bf 36}, 137 (1987).

\bibitem{prd89kxw}
Xian-Wei Kang, Bastian Kubis, Christoph Hanhart, and Ulf-G. Meiner
Phys. Rev. D {\bf 89}, 053015 (2014).
\bibitem{Daub:2015xja}
  J.~T.~Daub, C.~Hanhart and B.~Kubis,
  arXiv:1508.06841 [hep-ph].

\bibitem{xiao13}
C.W. Xiao and E. Oset,
Eur. Phys. J. A  {\bf 49}, 52 (2013).

\end{thebibliography}

\end{document}